\documentclass[article,accept,moreauthors,pdftex,10pt,a4paper,preprints]{mdpi}

\firstpage{1} 
\history{}

\usepackage[utf8]{inputenc}
\usepackage[T1]{fontenc}
\usepackage{hyperref}
\usepackage{graphicx,bm,amsmath,color,scalerel}
\usepackage{bbold}
\usepackage{wasysym}
\usepackage{verbatim}

\newcommand{\be}{\begin{equation}}
\newcommand{\ee}{\end{equation}}
\newcommand{\beq}{\begin{eqnarray}}
\newcommand{\eeq}{\end{eqnarray}}
\newcommand{\ba}{\begin{align}}
\newcommand{\ea}{\end{align}}

\newcommand{\bigo}{\,\scaleobj{1.3}{\oplus}\,} 

\Title{Observers and their notion of spacetime beyond special relativity}

\Author{José Manuel Carmona *\orcidA{}, José Luis Cortés \orcidB{} and José Javier Relancio \orcidC{}}

\AuthorNames{José Manuel Carmona, José Luis Cortés and José Javier Relancio}

\address[1]{%
Departamento de F\'{\i}sica Te\'orica,
Universidad de Zaragoza, Zaragoza 50009, Spain; jcarmona@unizar.es (J.M.C); cortes@unizar.es (J.L.C.); relancio@unizar.es (J.J.R.)}

\corres{Correspondence: jcarmona@unizar.es}

\abstract{It is plausible that quantum gravity effects may lead us to a description of Nature beyond the framework of special relativity. In this case, either the relativity principle is broken or it is maintained. These two scenarios (a violation or a deformation of special relativity) are very different, both conceptually and phenomenologically. We discuss some of their implications on the description of events for different observers and the notion of spacetime.}


\begin{document}

\section{Introduction}

\subsection{Beyond the framework of special relativity}

Poincaré invariance is at the root of our modern theories of particle physics. It is indeed the symmetry of the classical spacetime of special relativity (SR), which is one of their basic ingredients. These theories, however, are nowadays understood as low-energy limits of 
more fundamental, not yet known, constructions, where these essential ingredients might no longer be valid (see, for example, the different contributions in Ref.~\cite{Oriti:2009zz}). In particular, effects coming from a quantum theory of gravity (such as the creation and evaporation of virtual black holes~\cite{Kallosh:1995hi}) are expected to modify the classical spacetime picture, and therefore, also its symmetries. In this case, Poincaré invariance would then be a symmetry of spacetime in the low-energy limit. This reasoning makes plausible an intermediate regime where gravity and quantum effects can be neglected ($\hbar\to0$, $G\to 0$) but there is still a footprint of these effects ($M_P=\sqrt{\hbar c/G}\neq 0$, where $M_P$ is the Planck mass) producing a departure of SR. 

Once one accepts that SR may be an approximate symmetry, one can ask about the possibility that the scale $\Lambda$ controlling departures from SR be much lower than $M_P$, $\Lambda \ll M_P$. The phenomenological consistency of this possibility depends in fact on how one goes beyond the framework of special relativity, with two main options: either there is no a relativity principle, or there is one. In the first case, we speak of Lorentz invariance violation (LIV) \cite{Colladay:1998fq}, while in the second case, we speak of a deformed special relativity (or doubly special relativity, DSR~\cite{AmelinoCamelia:2008qg}).

Conceptually, LIV and DSR are very different. In LIV there is a special observer, that is, a privileged system of reference; in DSR, however, there is a class of equivalent observers (inertial reference systems), which are related through $\Lambda$-deformed Lorentz transformations. In the case of LIV, its dynamics is usually studied in the framework of local effective field theory (EFT), in which modified dispersion relations (MDR) for particles appear, thereby affecting the kinematics of SR; energy-momentum conservation laws are however unaltered in this context. This is however different in the case of DSR, since non-linear $\Lambda$-deformed Lorentz transformations are incompatible with linear energy-momentum conservation laws: the kinematics of DSR therefore contains MDR and MCL (modified composition laws); on the other hand, there is not a clear dynamical framework of DSR.

\subsection{Modified conservation laws and locality}
\label{sec:MCL-locality}

The conservation of the total momentum $\mathcal{P}$ of a system of particles is a reflection of an invariance under the transformations generated by $\mathcal{P}$ (translations). In the LIV case, and for the EFT framework,\footnote{Note that it is possible to consider LIV with a MCL, which goes beyond the EFT framework. In such a case, the MDR and the MCL would not be related by the compatibility relationship that the relativity principle establishes in DSR theories.} the total momentum is just the sum of the momenta of each of the particles, $\mathcal{P}=\sum p_I$, and translations are constant displacements, $x_I\to x_I+a$ $\forall I$. As we argued in the previous subsection, the total momentum of a system of particles in DSR is a non-linear composition of momenta, $\mathcal{P}=\bigo p_I$, and then translations (the transformations generated by $\mathcal{P}$) are momentum-dependent displacements, $x_I \to x_I+f(\{p\})$.

As a consequence of the above, LIV (in the EFT framework) has the property of \textbf{absolute} locality, as in SR: if two worldlines meet at a point, this will also happen for translated observer. In DSR, however, translations are not constant displacements, but they are momentum-dependent. Therefore, a local interaction for one observer is seen as non-local for a DSR-translated observer.

Nontrivial translations in DSR are in fact needed to avoid inconsistencies with tests of locality~\cite{Hossenfelder:2010tm,AmelinoCamelia:2010qv}. These inconsistencies derived from the hypothesis that crossing worldlines must also cross for a Lorentz-transformed observer. However, this hypothesis can be proved only in the case of absolute locality: in such a case, the crossing of two wordlines is an absolute fact for any observer, and in particular, for an observer whose origin coincides with the point of crossing. Since a Lorentz transformation does not affect the origin of coordinates and there exists absolute locality, the two lines will cross for any Lorentz-transformed observer. The failure of this hypothesis leading to inconsistencies with tests of locality, therefore, implies a loss of absolute locality.

On the other hand, if we define an interaction as the crossing of worldlines when the interaction point coincides with the origin of the observer, this is a relativistic invariant definition. Therefore, in DSR, which is a relativistic theory, all observers can define a spacetime region where interactions correspond to crossing of worldlines with a very good approximation. Far away from this region, however, interactions are seen as highly non-local. This property of DSR theories has been named as \emph{relative locality}~\cite{AmelinoCamelia:2011bm}. Note that relative locality was first described (\cite{AmelinoCamelia:2011bm}) as a property arising in a model of classical worldlines from a variational principle (we will review this construction in Sec.~\ref{sec:localints}), but, as we have shown here, the loss of absolute locality is a simple and generic consequence of a MCL.

The property of SR that an interaction corresponds to a crossing of worldlines is therefore no longer valid in DSR. This means that while, as we will see in Sec.~\ref{sec:localints}, interactions may still be used to define points in a spacetime, the structure of this spacetime in DSR will be more complicated than in SR or in LIV.

\subsection{Phenomenology in LIV and DSR}

The new high-energy effects (effects that are suppressed when $\Lambda\to\infty$) require astrophysical observations sensitive enough to the modifications to kinematics and/or cumulative effects, such as long distance propagations; the phenomenology, however, is radically different in the LIV and DSR scenarios.

LIV affects strongly thresholds of reactions (with corrections of order $E^3/m^2\Lambda$) and is able to forbid/allow decays at high energy that are allowed/forbidden by SR, what can then produce observable spectrum changes after a long propagation. In contrast, DSR cannot change the low-energy (allowed or forbidden) character of a decaying particle at high energy because of the relativity principle, and reaction thresholds are also much less changed because of cancellations between the modifications in the dispersion relation and the composition laws (essentially, changes are described in terms of the relativistic invariant of the reaction, not in terms of one of the energies)~\cite{Carmona:2014aba}. 

DSR phenomenology is therefore much more limited that LIV phenomenology, and only involves amplification mechanisms in the propagation of stable particles: photons and neutrinos. Since vacuum birrefringence is not a generic feature in DSR, this leaves time of flight as the only window to leading order modifications of SR which can be compatible with a relativity principle. The analysis of time of flight in the LIV and the DSR cases is however very different: in the LIV case, it only depends on the particle dispersion relation; in the DSR case, it also depends on the spacetime structure.

In this work, we will review a recently proposed~\cite{Carmona:2017cry} model for the nontrivial spacetime structure that arises in
the description of physical measurements made by translated observers in DSR theories, and in view of this result, we will update the time delay calculation presented in Ref.~\cite{Carmona:2017oit}, giving a new perspective on the problem.

\section{Spacetime through local interactions}
\label{sec:localints}

\subsection{Relative locality and a first attempt to recover absolute locality in a new spacetime}

Local interactions in field theory are associated to linear momentum composition laws, since the product of fields at a single point gives a $\delta(\sum p_i)$ in momentum space. Nonlinear momentum composition laws will be, therefore, associated with a change in the concept of locality, and lie beyond the EFT framework. Indeed, we have argued that a MCL of momenta is associated with a loss of the notion of absolute (observer-independent) locality of interactions.
Although there is no at present a consistent field theory treatment of DSR theories, the loss of absolute spacetime locality can be easily incorporated in a classical model of worldlines through a variational principle: this is the \textbf{relative locality} framework~\cite{AmelinoCamelia:2011bm}.

Take $N$ interacting particles, numbering from 1 to $N$ the incoming worldlines, and from $N+1$ to $2N$ the outgoing worldlines:
\begin{align}
S_\text{total}&= S_\text{free}^\text{in}+S_\text{free}^\text{out} +S_\text{int} \nonumber \\
S_\text{free}^\text{in}&=\sum_{J=1}^{N}\int^{0}_{-\infty} d\tau \left(x^{\mu}_J \dot p^{J}_{\mu}+\mathcal{N}_J\left(C(p^{J})-m^2_J\right)\right) \nonumber \\
S_\text{free}^\text{out}&=\sum_{J=N+1}^{2N}\int_{0}^{\infty} d\tau \left(x^{\mu}_J \dot p^{J}_{\mu}+\mathcal{N}_J\left(C(p^{J})-m^2_J\right)\right) \nonumber \\
S_\text{int}&=\left(\underset{N+1\leq J\leq 2N}{\bigoplus} p^J_\nu(0)\,\,-\underset{1\leq J\leq N}{\bigoplus} p^J_\nu(0)\right) \xi^\nu \,,
\label{eq:action}
\end{align}
where the $\mathcal{N}_J$ are Lagrange multipliers that ensure the satisfaction of the modified dispersion relation $C(p^J)=m_J^2$.
The parameter $\tau$ along each worldline has been chosen such that $\tau=0$ corresponds to the interaction. The $2N$ worldlines are determined by the condition $\delta S_\text{total}=0$ for any variation $\delta\xi^\mu$, $\delta x_J^\mu$, $\delta p^J_\mu$. From the variation  $\delta p_\mu^J(0)$ one finds 
\begin{equation}
x^{\mu}_J (0)\,=\, \xi^{\nu} \frac{\partial \mathcal{P}_\nu}{\partial p^J_{\mu}(0)} \,\forall J \,,
\label{eq:vertex}
\end{equation}
and the variation $\delta\xi^\mu$ leads to identify the total momentum $\mathcal{P}=\underset{1\leq J\leq N}{\bigoplus} p^J(0)=\underset{N+1\leq J\leq 2N}{\bigoplus} p^J(0)$, conserved in the interaction.

From Eq.~\eqref{eq:vertex} we get absolute locality, $x^{\mu}_J (0)\,=\, \xi^{\mu}$, if and only if $\mathcal{P}$ is the linear addition of momenta. We can ask then the following question: can one define a new spacetime $\tilde{x}$ from the canonical spacetime $(x,p)$,  $\{p_\nu,\tilde{x}^\mu\}=\delta^\mu_\nu$, such that we recover absolute locality in this new spacetime?

As a first attempt, let us introduce a nontrivial spacetime $(\tilde{x})$ through the definition
\begin{equation}
\tilde{x}^\mu \,=\, x^\nu \, \varphi^\mu_\nu(p/\Lambda),
\label{eq:NCdef}
\end{equation}
where $\varphi^\mu_\nu$ is such that $\varphi^\mu_\nu(0)=\delta^\mu_\nu$. This new spacetime is `noncommutative', in the sense
\begin{equation}
\lbrace \tilde{x}^\mu, \tilde{x}^\nu\rbrace \,=\, -x^\sigma \varphi^\mu_\rho \frac{\partial\varphi^\nu_\sigma}{\partial p_\rho} + x^\rho \varphi^\nu_\sigma \frac{\partial\varphi^\mu_\rho}{\partial p_\sigma},
\label{eq:NC}
\end{equation}
and the Poisson brackets in the new phase space are now: $\{p_\nu,\tilde{x}^\mu\}=\varphi^\mu_\nu$.

Now let us consider a process in which we have a particle in the initial state with momentum $k$, and two particles in the final state with momenta $p$ and $q$, such that $k=p\oplus q$. Since the coordinates of the two outgoing worldlines at the interaction are
\begin{equation}
y^\mu(0) \,=\, \xi^\nu \frac{\partial(p\oplus q)_\nu}{\partial p_\mu} \,,\quad z^\mu(0) \,=\, \xi^\nu \frac{\partial(p\oplus q)_\nu}{\partial q_\mu},
\label{eq:y-z-vertex}
\end{equation}
and the coordinate of the first particle is $x^\mu(0)=\xi^\mu$, the condition to have a local interaction in the noncommutative spacetime is
\begin{equation}
\boxed{\varphi^\mu_\nu(p\oplus q) \,=\, \frac{\partial (p\oplus q)_\nu}{\partial p_\rho} \,\varphi^\mu_\rho(p) \,=\, \frac{\partial (p\oplus q)_\nu}{\partial q_\rho} \, \varphi^\mu_\rho(q),}
\label{eq:condition-1}
\end{equation}
where we have imposed a common interaction point
\begin{equation}
\tilde{x}^\mu(0)=\xi^\nu \varphi^\mu_\nu(p\oplus q)=\tilde{y}^\mu(0)=\tilde{z}^\mu(0).
\label{eq:commonNC}
\end{equation}

However, taking the limits $p\to 0$ and $q\to 0$ in Eq.~\eqref{eq:condition-1}, one gets
\begin{equation}
\varphi^\mu_\nu(p)=\lim_{q\to 0} \frac{\partial (q\oplus p)_\nu}{\partial q_\mu} \,=\,\lim_{q\to 0} \frac{\partial (p\oplus q)_\nu}{\partial q_\mu}
\label{eq:comm}
\end{equation}
which is valid in the case of a commutative MCL, but not for more general cases. Moreover, in this case the $\tilde{x}$ are commutative coordinates, and can then be seen as just a choice of spacetime coordinates in a canonical phase space
\cite{Carmona:2017cry}.

\subsection{A second attempt to define spacetime in the two-particle system}

Since in relative locality the amount of non-locality depends on the total momentum of the process, it is natural to consider that the noncommutative spacetime in which the interaction is local should depend on the momenta of the rest of worldlines. For the two out-going particles,
\begin{equation}
\tilde{y}^\mu \,=\, y^\nu \,\varphi_{L\,\nu}^{\,\mu}(p, q) \,,\quad
\tilde{z}^\mu \,=\, z^\nu \, \varphi_{R\,\nu}^{\,\mu}(p, q).
\label{eq:y-z-NC}
\end{equation}
The condition to have an event defined by the interaction is then
\begin{equation}
\boxed{\varphi^\mu_\nu(p\oplus q) \,=\, \frac{\partial (p\oplus q)_\nu}{\partial p_\rho} \,\varphi_{L\,\rho}^{\,\mu}(p, q) \,=\, \frac{\partial (p\oplus q)_\nu}{\partial q_\rho} \, \varphi_{R\,\rho}^{\,\mu}(p, q).} 
\label{eq:condition-2}
\end{equation}

Taking again the limits $p\to 0$ and $q\to 0$ in Eq.~\eqref{eq:condition-2}, one obtains
\begin{equation}
\varphi_{L\,\sigma}^{\:\:\mu}(p, 0) = \varphi^\mu_\sigma(p)\,, \quad \varphi_{R\,\sigma}^{\:\:\mu}(0, q) = \varphi^\mu_\sigma(q).
\label{eq:bordercond}
\end{equation}
However, the MCL does not unequivocally determine in this case the functions $\varphi,\varphi_L,\varphi_R$.

A specially simple case is if one takes $\varphi^{\:\:\mu}_{L\rho}$ to depend only on one momentum; then one has
$\varphi^{\:\:\mu}_{L\rho}(p, q) \,=\, \varphi^{\:\:\mu}_{L\rho}(p, 0) \,=\,\varphi^\mu_\rho(p)$, and 
\begin{equation}
\varphi^\mu_\nu(p\oplus q) \,=\, \frac{\partial (p\oplus q)_\nu}{\partial p_\rho} \,\varphi^\mu_\rho(p),
\label{eq:MCL-NC}
\end{equation}
which can be used to determine the MCL for a given one-particle noncommutative spacetime $\varphi$. This case gives, in particular,
\begin{equation}
\varphi^\mu_\nu(p)  \,=\, \lim_{q\to 0}  \frac{\partial (q\oplus p)_\nu}{\partial q_\mu},
\label{eq:master-eq}
\end{equation}
which has a simple interpretation: the transformation generated by the noncommutative spacetime coordinates is a displacement in momentum space defined by the MCL:
\begin{equation}
\delta p_\mu \,=\, -\epsilon_\nu \lbrace\tilde{x}^\nu, p_\mu\rbrace \,=\, \epsilon_\nu \varphi^\nu_\mu(p) \,=\, \epsilon_\nu \lim_{q\to 0} \frac{\partial(q\oplus p)_\mu}{\partial q_\nu} \,=\, \left[(\epsilon\oplus p)_\mu - p_\mu\right].
\label{eq:generator}
\end{equation}

\subsection{Application to $\kappa$-Poincaré}

The Hopf algebra $\kappa$-Poincaré offers also a link between a MCL (that can be read from the coproduct of the algebra) and a noncommutative spacetime (in this case, $\kappa$-Minkowski). We can show that this link is exactly the one described above.

In the bicrossproduct basis of $\kappa$-Poincaré~\cite{KowalskiGlikman:2002jr}, the coproduct of the generators of translations $P_\mu$ reads
\begin{equation}
\Delta(P_0)\,=\,P_0\otimes \mathbb{I}+ \mathbb{I}\otimes P_0 \,,\qquad \Delta(P_i)\,=\,P_i\otimes \mathbb{I}  + e^{-P_0/\Lambda} \otimes P_i,
\label{eq:coprod}
\end{equation}
which defines the momentum composition law
\begin{equation}
(p\oplus q)_0\,=\,p_0 + q_0 \,, \qquad (p\oplus q)_i \,=\,p_i  + e^{-p_0/\Lambda} q_i.
\label{eq:MCLbicr}
\end{equation}

Phase-space Poisson brackets are obtained through the method of `pairing'~\cite{Kosinski_paring}, which gives
\begin{equation}
\lbrace\tilde{x}^0, \tilde{x}^i\rbrace \,=\,-\frac{\tilde{x}^i}{\Lambda}\,,\qquad  \lbrace\tilde{x}^0, p_0\rbrace \,=\,-1\,,\qquad  \lbrace\tilde{x}^0, p_i\rbrace \,=\,\frac{p_i}{\Lambda}\,,\qquad  \lbrace\tilde{x}^i, p_j\rbrace \,=\,-\delta^i_j\,,
\qquad \lbrace\tilde{x}^i, p_0\rbrace \,=\,0,
\label{eq:PP}
\end{equation}
or, equivalently,
\begin{equation}
\varphi^0_0(p)=1 \,,\qquad \varphi^0_i(p)=-\frac{p_i}{\Lambda} \,,\qquad \varphi^i_j(p)=\delta^i_j \,,\qquad \varphi^i_0(p)=0.
\label{eq:phi}
\end{equation}

The functions $\varphi^\mu_\nu(p)$ and the composition law $(p\oplus q)$ above are indeed related by Eq.~\eqref{eq:MCL-NC}.	The locality condition gives then a physical interpretation of the `pairing' procedure that introduces spacetime in $\kappa$-Poincaré in a specific way.

From this example, $\kappa$-Minkowski spacetime can be seen as the (one-particle) noncommutative spacetime that emerges from a locality condition in a classical model which generalizes SR in momentum space through the algebra of $\kappa$-Poincaré. This is a rather different approach from the introduction of noncommutativity as the implementation of a possible minimal length in a quantum spacetime.

Note that in the general case, the implementation of locality is compatible with an independent choice for a one-particle noncommutative spacetime (the $\varphi$ function) and a MCL, while these two ingredients are related in the previous example. From this perspective, then, the example of $\kappa$-Poincaré Hopf algebra is a very particular case of the implementation of locality leading to a relativistic generalization of SR. In this case, the two particles coming out of the interaction propagate in (different) noncommutative spacetimes that are defined by the functions $\varphi_{L\,\nu}^{\:\:\mu}(p, q)\,=\, \varphi^\mu_\nu(p)$ and $\varphi_{R\,\nu}^{\:\:\mu}(p, q)$,
\begin{equation}
\varphi_{R\,0}^{\:\:0}(p, q)\,=\,1 \,,\qquad \varphi_{R\,0}^{\:\:i}(p, q)\,=\,0 \,,\qquad \varphi_{R\,i}^{\:\:0}(p, q)\,=\,-\frac{e^{p_0/\Lambda}p_i+q_i}{\Lambda}\,, \qquad \varphi_{R\,j}^{\:\:i}(p, q)\,=\,e^{p_0/\Lambda}\delta^i_j .
\label{eq:phiRbicr}
\end{equation}
From this, one obtains the two-particle spacetime Poisson brackets that are different from zero:
\begin{equation}
\lbrace\tilde{y}^0, \tilde{y}^i\rbrace \,=\,-\frac{\tilde{y}^i}{\Lambda}\,, \qquad
\lbrace\tilde{z}^0, \tilde{z}^i\rbrace \,=\,-\frac{\tilde{z}^i}{\Lambda}\,,\qquad \lbrace\tilde{y}^0, \tilde{z}^i\rbrace \,=\,
\lbrace\tilde{z}^0, \tilde{y}^i\rbrace \,=\,
-\frac{\tilde{z}^i}{\Lambda}.
\label{eq:PPbicr}
\end{equation}


\subsection{More than two particles}

We have obtained the locality condition Eq.~\eqref{eq:condition-2} for an interaction with two particles in the final state. It is possible, however, to generalize this condition to an arbitrary number of particles. As an example, let us consider an initial state where all except one of the momenta are arbitrarily small and a final state of three particles with momenta $(k, p, q)$ and a total momentum ${\cal P} = (k\oplus p)\oplus q$. Then one has 
\be
X^\rho(0) \,=\, \xi^\rho  \,, {\hskip 0.6cm} 
x^\nu(0) \,=\, \xi^\rho \frac{\partial {\cal P}_\rho}{\partial k_\nu} \,, {\hskip 0.6cm} 
y^\nu(0) \,=\, \xi^\rho \frac{\partial {\cal P}_\rho}{\partial p_\nu} \,, {\hskip 0.6cm} 
z^\mu(0) \,=\, \xi^\rho  \frac{\partial {\cal P}_\rho}{\partial q_\nu} \,.  
\ee
We can introduce new spacetime coordinates 
\be
\tilde{X}^\mu \,=\, X^\nu \varphi^\mu_\nu({\cal P}) \,,{\hskip 0.6cm} 
\tilde{x}^\mu \,=\, x^\nu \varphi^{\:\mu}_{1\,\nu}(k, p, q) \,,{\hskip 0.6cm}
\tilde{y}^\mu \,=\, y^\nu \varphi^{\:\mu}_{2\,\nu}(k, p, q) \,,{\hskip 0.6cm}
\tilde{z}^\mu \,=\, z^\nu \varphi^{\:\mu}_{3\,\nu}(k, p, q) \,.
\ee
If we choose the functions $\varphi_i$ such that
\be
\varphi^\mu_\rho({\cal P}) \,=\, \frac{\partial {\cal P}_\rho}{\partial k_\nu}  \varphi^{\:\mu}_{1\,\nu}(k, p, q) \,=\, \frac{\partial {\cal P}_\rho}{\partial p_\nu}  \varphi^{\:\mu}_{2\,\nu}(k, p, q) \,=\, \frac{\partial {\cal P}_\rho}{\partial q_\nu}  \varphi^{\:\mu}_{3\,\nu}(k, p, q) \,,
\label{loc3}
\ee
then one has all the particles at the same point in the interaction 
\be
\tilde{X}^\mu(0) \,=\, \tilde{x}^\mu(0) \,=\, \tilde{y}^\mu(0) \,=\, \tilde{z}^\mu(0) \,,
\ee
and locality is implemented in the new spacetime. Using now the chain rule for the partial derivatives:
\be
\frac{\partial {\cal P}_\rho}{\partial k_\nu} \,=\, \frac{\partial {\cal P}_\rho}{\partial(k\oplus p)_\sigma} \frac{\partial(k\oplus p)_\sigma}{\partial k_\nu} \,{\hskip 0.7cm}
\frac{\partial {\cal P}_\rho}{\partial p_\nu} \,=\, \frac{\partial {\cal P}_\rho}{\partial(k\oplus p)_\sigma} \frac{\partial(k\oplus p)_\sigma}{\partial p_\nu} \,,
\ee
one has
\be
\varphi^\mu_\rho({\cal P}) \,=\, \frac{\partial {\cal P}_\rho}{\partial(k\oplus p)_\sigma} \frac{\partial(k\oplus p)_\sigma}{\partial k_\nu} \varphi^{\:\mu}_{1\,\nu}(k, p, q) \,=\, 
\frac{\partial {\cal P}_\rho}{\partial(k\oplus p)_\sigma} \frac{\partial(k\oplus p)_\sigma}{\partial p_\nu} \varphi^{\:\mu}_{2\,\nu}(k, p, q) \,=\, \frac{\partial {\cal P}_\rho}{\partial q_\nu}  \varphi^{\:\mu}_{3\,\nu}(k, p, q) \,.
\ee
This requires that
\be
\frac{\partial(k\oplus p)_\sigma}{\partial k_\nu} \varphi^{\:\mu}_{1\,\nu}(k, p, q) \,=\, 
\frac{\partial(k\oplus p)_\sigma}{\partial p_\nu} \varphi^{\:\mu}_{2\,\nu}(k, p, q) \,=\, \varphi_{L\,\sigma}^{\:\mu}((k\oplus p), q) \,,{\hskip 0.7cm}  \varphi^{\:\mu}_{3\,\nu}(k, p, q) \,=\, \varphi_{R\,\sigma}^{\:\mu}((k\oplus p), q)\,,
 \label{varphi(1-2-3)}
\ee
which can be used to determine $\varphi_1$, $\varphi_2$ and $\varphi_3$. 

Therefore, a given composition law of momenta ($p\oplus q$) and a noncommutative one-particle spacetime ($\varphi$) allow one to determine the spacetime of a two particle system (the fuctions $\varphi_L$, $\varphi_R$), and Eqs.~\eqref{varphi(1-2-3)} give the spacetime of a three particle system implementing locality. 


\section{Phenomenology in the relative locality framework}
\label{sec:NC}

\subsection{Two notions of spacetime}

In the previous section we have shown that, even in the framework of relative locality, where worldlines of particles do not generically cross in the $x$-spacetime (where $x$ is canonically conjugated to the momentum variable $p$), interactions may serve to define a notion of spacetime. Indeed, each interaction has a set of four coordinates $\xi^\mu$ associated to it, as defined in the action Eq.~\eqref{eq:action}. These coordinates might serve as a definition for spacetime. Thus, if observer $A$ sees an interaction ocurring ``at'' coordinates $\xi^\mu_A$, a translated observer (with parameters of translation $a^\mu$) will associate that interaction with coordinates 
\begin{equation}
\xi^\mu_B=\xi^\mu_A+a^\mu\,.
\label{eq:xi}
\end{equation}

The description of worldlines is however not simple in such a spacetime, because outgoing (incoming) worldlines begin (end) at different points of that spacetime, which depend on the whole set of momenta that participate in the interaction. The use of the noncommutative spacetime introduced in the previous subsections makes possible to have a common interaction point of coordinates $\zeta^\mu$, whose relation with the $\xi^\mu$ coordinates is [see Eq.~\eqref{eq:commonNC}]:
\begin{equation}
\zeta^\mu=\xi^\nu \varphi^\mu_\nu(\mathcal{P}/\Lambda),
\label{eq:zeta-xi}
\end{equation}
where $\mathcal{P}$ is the total momentum of the interaction. 

Coordinates $\zeta^\mu$ might serve as well as coordinates $\xi^\mu$ to define a notion of spacetime, and in this case they have the advantage to maintain absolute locality for interactions, that is: every observer agrees that wordlines cross at interactions. In such a spacetime, the coordinates of an interaction point for observer $A$, $\zeta^\mu_A$, are related to the coordinates of a translated observer $B$ (with parameters of translation $b^\mu$) by
\begin{equation}
\zeta^\mu_B=\zeta^\mu_A+b^\mu\,.
\label{eq:zeta}
\end{equation}

Compatibility of Eqs.~\eqref{eq:xi} and~\eqref{eq:zeta} requires, according to Eq.~\eqref{eq:zeta-xi}, the relation
\begin{equation}
  a^\nu \varphi^\mu_\nu(\mathcal{P}/\Lambda) = b^\mu
\end{equation}
between the parameters of the translation in the two spacetimes. Then, if one identifies the parameters of a translation with a fixed distance between the two observers, it is clear that the two notions of spacetime defined above are different alternatives that have in fact different phenomenological consequences.

\subsection{Photon time delays in a DSR theory: a first possibility}

Any physical measurement involves at least two interactions: in the simplest case, the emission of a photon by a (static) source and the detection of this photon by a (static) detector. Two observers for which each of the interactions is respectively local in the canonical spacetime, can then be brought into play to describe this measurement: observer $A$, at the source, and observer $B$, at the detector.

Let us first consider a model of spacetime that derives directly from the noncommutativity, that is, the model that is defined by Eq.~\eqref{eq:zeta}. When we say that source and detector are in relative rest and separated by a distance $L$, we have that
\begin{equation}
\vec{\zeta}_B=\vec{\zeta}_A-(0,0,L),
\label{eq:zetaTD}
\end{equation}
where we have taken that the propagation of the photon takes place along the third spatial direction. 

Let us now consider the detection of a photon which is produced by a certain source in some more or less complicated process (for example, a gamma-ray burst). Detection of this photon takes place for observer $B$ at $\vec{\tilde{x}}_d=(0,0,0)$ and is emitted at $\vec{\tilde{x}}_{e}=(0,0,-L)$, which corresponds to the origin of spatial coordinates for observer $A$, according to Eq.~\eqref{eq:zetaTD}.

The time coordinates at the detection and emission of the photon will be related by
\begin{equation}
\tilde{x}^0_{d}=\tilde{x}^0_{e}+\frac{L}{\tilde{v}}\,,
\label{eq:TD1}
\end{equation}
where $\tilde{v}$ is the velocity of propagation of the photon. Using the expression of the space-time coordinates in terms of canonical coordinates
\begin{equation}
\tilde{x}^\mu=x^\nu\varphi^\mu_{\nu}(\{p\}),
\label{eq:NC-C}
\end{equation}
where $(\{p\})$ represents the set of momenta involved in the interaction (as in Eq.~\eqref{eq:y-z-NC}), one gets
\begin{equation}
\tilde{v}(\{p\})=\frac{(x^0_{d}-x^0_{e})\varphi^3_{0}+(x^3_{d}-x^3_{e})\varphi^3_{3}}{(x^0_{d}-x^0_{e})\varphi^0_{0}+(x^3_{d}-x^3_{e})\varphi^0_{3}}=\frac{\varphi^3_{0}+v(p)\varphi^3_{3}}{\varphi^0_{0}+v(p)\varphi^0_{3}}\,,
\label{eq:NCv-Cv}
\end{equation}
where $v(p)$ is the velocity in the canonical spacetime, which can be derived from the action Eq.~\eqref{eq:action}:
\begin{equation}
v(p)=\frac{\partial C(p)/\partial p_3}{\partial C(p)/\partial p_0}\,.
\label{eq:Cv}
\end{equation}

A difficulty remains: the fact that $\varphi^\mu_{\nu}(\{p\})$ could depend on all the momenta involved in the interaction would make impossible to calculate in practice the time interval between the emission and detection of the photon. However, we have seen that it is possible to implement the locality condition Eq.~\eqref{eq:condition-2} in such a way that the function  $\varphi^\mu_{\nu}$ for one of the particles depends only on its own momentum $p$ (this choice gave the relationship between the MCL and the one-particle noncommutative spacetime Eq.~\eqref{eq:MCL-NC}). If we adopt this choice for the emitted photon which will arrive to the detector, $\tilde{v}$ is a function of $p$ only, and can then be calculated for a given noncommutative spacetime $\varphi^\mu_\nu(p)$.
 
Within this scheme, Eq.~\eqref{eq:TD1} gives a time delay $\tilde{T}$ 
\begin{equation}
\tilde{T}\doteq \tilde{x}^0_d - \tilde{x}^0_e - L =L\left(\frac{1}{\tilde{v}}-1\right)\,.
\label{eq:TD-2}
\end{equation}

As a particular example, one can compute $\tilde{v}$ in the bicrossproduct basis of $\kappa$-Poincaré using Eq.~\eqref{eq:phi} and the velocity in the canonical spacetime Eq.~\eqref{eq:Cv} determined by the Casimir in this basis~\cite{KowalskiGlikman:2002jr}. The result is $\tilde{v}=1$, and, therefore, there is no time delay in this basis.
In fact, the different bases of $\kappa$-Poincaré correspond to different choices of canonical phase-space coordinates with the same noncommutative spacetime~\cite{Carmona:2017cry}; then, the absence of time delay is a property of $\kappa$-Poincaré, independent of the choice of basis.

In Ref.~\cite{Carmona:2017oit} a different expression for a `time delay' was obtained. It used in fact a different definition of the distance $L$ between the detector and the source (difference of canonical space coordinates instead of noncommutative space coordinates). The absence of time delay was in that case a consequence of a velocity of propagation in canonical spacetime independent of energy ($v=1$), which is a 
basis-dependent property, and, in particular, one has a time delay in the bicrossproduct basis.

\subsection{Photon time delays in a DSR theory: a second possibility}

Let us now consider the second model of spacetime, in which space-time coordinates are defined with the vertex coordinates $\xi^\mu$ of an interaction, Eq.~\eqref{eq:vertex}. One can follow step by step the definition of the distance $L$ between source and detector in relative rest in the previous section, replacing everywhere the coordinates $\tilde{x}_d^\mu$ by $\xi_d^\mu$ and $\tilde{x}_e^\mu$ by $\xi_e^\mu$.

The computation of the time delay $T\doteq \xi^0_d-\xi^0_e-L$ requires the relations
\begin{equation}
  x^\mu_d=\xi^\nu_d \frac{\partial \mathcal{P}_\nu}{\partial p_\mu} \quad \quad
  x^\mu_e=\xi^\nu_e \frac{\partial \mathcal{P}_\nu}{\partial p_\mu}\,,
\end{equation}
and the equation of the worldline in canonical spacetime corresponding to the velocity Eq.~\eqref{eq:Cv}. The result is
\begin{equation}
  T=L\left[\frac{(\partial C/\partial p_0)(\partial \mathcal{P}_3/\partial p_3)-(\partial C/\partial p_3)(\partial \mathcal{P}_3/\partial p_0)}{(\partial C/\partial p_3)(\partial \mathcal{P}_0/\partial p_0)-(\partial C/\partial p_0)(\partial \mathcal{P}_0/\partial p_3)}-1\right].
  \end{equation}

The derivatives $\partial\mathcal{P}_\nu/\partial p_\mu$ appearing in the previous expression inevitably depend on all the momenta involved in the process, making impossible to determine the time delay in terms of the energy of the photon. In particular, it is not possible to have zero time delay, in contrast with the previous choice of spacetime.

\section{Conclusions}

The phenomenology beyond SR is very different in the cases of absence of a relativity principle (LIV) and presence of a relativity principle (DSR). In contrast to a number of possible footprints in the LIV case, time of flight of stable particles may be the only phenomenological window in the DSR case. Moreover, the phenomenological analysis of time of flight is also different for the LIV and DSR cases.

The concept of absolute locality in canonical spacetime, which is valid in SR, is only possible in a LIV framework. DSR necessarily requires a modification in the notion of locality, that can be incorporated through a new (noncommutative) spacetime defined from the canonical phase space.

A definition of spacetime from interactions in the case of DSR requires to go beyond the canonical spacetime. The modification of the composition of momenta leads to a distribution of the canonical spacetime coordinates of the different particles in the interaction, which are fixed in terms of a vertex.

We have presented two options for the spacetime in DSR. One possibility is to consider new (noncommutative) space-time coordinates for each particle, such that the interaction is local in this new spacetime. A second possibility is to consider directly the vertex coordinates as the coordinates of spacetime.

In both cases one has in general a time delay depending on all the momenta in the process, and then one does not have a definite prediction. But in the case of a noncommutative spacetime we have shown that it is possible to find a model where one does not have time delays. This opens up the possibility to consider a scale in DSR much smaller than the Planck scale, leading to a new perspective in quantum gravity phenomenology.

Finally, we note that it is plausible that the recovery of locality in the new spacetime might serve as a starting point for a generalization of field theory based on modified composition laws. 

\acknowledgments{This work is supported by the Spanish MINECO FPA2015-65745-P (MINECO/FEDER) and Spanish DGIID-DGA Grant No. 2015-E24/2. We thank Flavio Mercati for the organization of the conference \emph{Observers in Quantum Gravity}, that took place in Rome from on 22-23 January 2018. This paper is based on a talk given at the mentioned conference.}

\abbreviations{The following abbreviations are used in this manuscript:\\

\noindent 
\begin{tabular}{@{}ll}
SR & Special Relativity \\
LIV & Lorentz Invariant Violation\\
DSR & Deformed Special Relativity\\
EFT & Effective Field Theory \\
MDR & Modified Dispersion Relation\\
MCL & Modified Composition Law
\end{tabular}}

\end{document}